\documentclass[twocolumn,letter,dvipdfmx]{jpsj3}
\usepackage{bm}
\usepackage{color}
\usepackage{ulem}
\usepackage{txfonts}

\title{
Distributed Hybridization Model for Quantum Critical Behavior\\ in Magnetic Quasicrystals
}

\author{%
Junya Otsuki$^1$ and
Hiroaki Kusunose$^2$
}

\inst{%
$^1$Department of Physics, Tohoku University, Sendai 980-8578, Japan\\
$^2$Department of Physics, Meiji University, Kawasaki 214-8571, Japan
}

\recdate{\today}

\abst{%
A quantum critical behavior of the magnetic susceptibility was observed in a quasicrystal containing ytterbium.
At the same time, a mixed-valence feature of Yb ions was reported, which appears to be incompatible with the magnetic instability.
We derive the magnetic susceptibility by expressing the quasiperiodicity as the distributed
hybridization strength between Yb $4f$ and conduction electrons. 
Assuming a wide distribution of the hybridization strength, the most $f$ electrons behave as renormalized
paramagnetic states in
the Kondo or mixed-valence regime,
but a small number of $f$ moments remain unscreened. As a result, the bulk magnetic susceptibility exhibits a nontrivial power-law-like behavior, while the average $f$-electron occupation is that of mixed-valence systems.
This model thus resolves
two contradictory properties of Yb quasicrystals.
}

\begin{document}
\maketitle

Quasicrystals, which were discovered in 1984 by Shechtman {\it et al.}~\cite{Shechtman84}, constitute a unique class of crystals.
Because of the absence of translational symmetry, the Bloch theorem is not applicable.
Theoretical investigations on their electronic properties have so far revealed, for example, the existence of a universal pseudogap from electronic structure calculations~\cite{Fujiwara91} and the emergence of confined states from model calculations~\cite{Kohmoto86,tsunetsugu86,Fujiwara88}.

Interesting magnetic properties were recently found by Deguchi {\it et al.} in a Tsai-type quasicrystal with ytterbium atoms Au$_{51}$Al$_{34}$Yb$_{15}$~\cite{Deguchi12}.
The susceptibility at $T \gtrsim 100$ K satisfies the Curie law with effective magnetic moment $\mu_{\rm eff} \approx 3.9 \mu_{\rm B}$, indicating a major contribution from the Yb$^{3+}$ ions with the $4f^{13}$ configurations.
The susceptibility $\chi$ continues to increase as $\chi\propto T^{-\gamma}$ with $\gamma \approx 0.5$ down to $T=0.1$ K,
but no phase transition has been observed.
The specific heat $C$ also exhibits an anomalous $T$-dependence, $C/T \sim -\log T$.
Interestingly, no divergence of $\chi$ and $C/T$ has been observed in the approximant crystal Au$_{51}$Al$_{35}$Yb$_{14}$, which has the same local structure but with periodicity. 
This strongly indicates that the lack of periodicity plays a key role in the observed ``quantum critical'' behaviors.
Furthermore, the robustness of the low-temperature properties against external pressure confirms the distinction from the ordinary quantum critical phenomena due to magnetic long-range ordering.
Motivated by these observations, the correlation effects in quasiperiodic lattices were investigated theoretically~\cite{Watanabe13,Takemori14,Takemura15,Shinzaki16}.

The $f$-electron valence of Yb ions was determined by X-ray absorption measurement~\cite{Watanuki12, Matsukawa14}. A mean valence of 2.61 was reported, meaning that magnetic Yb$^{3+}$ ions having the $4f^{13}$ configuration and nonmagnetic Yb$^{2+}$ ions are mixed.
The mixed-valence state is, in a naive picture, incompatible with the magnetic anomaly.
Discussion in terms of valence fluctuations has been invoked to connect the mixed-valence state with the anomalous magnetic properties\cite{Watanabe16}, although no direct evidence of valence fluctuations has been reported yet.

\begin{figure}[b]
	\begin{center}
	\includegraphics[width=0.85\linewidth]{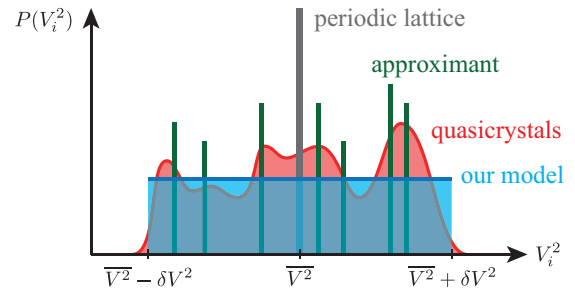}
	\end{center}
	\caption{(Color online) Schematic picture of the different distributions $P(V_i^2)$ of the hybridization strength $V_i^2$. }
	\label{fig:hyb_prob}
\end{figure}

In this Letter, we address the contradictory magnetic and mixed-valence properties of the Yb quasicrystal from another viewpoint, namely, the Kondo screening in quasicrystals without periodicity.
Strictly speaking, there are no equivalent sites in a crystallographic sense.
This means that the environment of $4f$ electrons in Yb ions, such as the number of neighboring Au/Al atoms and the distances to them,
is different from site to site.
This situation may be described by site-dependent local parameters, such as the hybridization strength $V_i$ between $f$ and conduction electrons.
In this perspective, the distribution of $V_i$ (or $V_i^2$) is the key quantity that distinguishes quasicrystals from ordinary periodic materials as well as their approximants.
We expect a continuous distribution in quasicrystals, while it consists of only a single or finite number of delta functions in periodic and approximant crystals.
Figure~\ref{fig:hyb_prob} schematically shows the various kinds of distributions of $V_{i}^{2}$.

If the spatial arrangement of $V_i$ is neglected, electronic properties are essentially determined by only the distribution function $P(V_i^2)$.
More specifically, $V_i^2$ may be randomly distributed in accordance with the probability distribution $P(V_i^2)$.
Such a model is known as the Kondo disorder model, which was discussed in the context of the ``quantum critical'' behaviors observed in heavy-fermion alloys with considerable disorder~\cite{Bernal95,Miranda96,Miranda97}.
The above circumstances suggest that the Kondo disorder scenario is also applicable
to quasicrystals {\it without} disorder.
In other words, if relevant electrons are subjected to essentially {\it local} environments, which differ from site to site, in quasicrystals, quasiperiodicity cannot be distinguished from randomly distributed local environments.
This scenario was recently proposed by Andrade {\it et al.}\cite{Andrade15}
They computed the site dependence of the hybridization strength on a model quasiperiodic lattice~\cite{footnote-Andrade} and
demonstrated correspondence with the Kondo disorder picture.
Nevertheless, further numerical investigations are required to elucidate the valence properties as well as the explicit temperature variation of physical quantities of interest. 
This is the aim of this Letter.

We consider an Anderson lattice model with site-dependent hybridization.
Using a hole picture, we represent the $4f^{13}$ ($4f^{14}$) configuration of Yb$^{3+}$ (Yb$^{2+}$) ions as the $f^1$ ($f^0$) state. The Hamiltonian reads
\begin{align}
{\cal H} &= \sum_{\bm{k}\alpha} (\epsilon_{\bm{k}} - \mu) c_{\bm{k}\alpha}^{\dag} c_{\bm{k}\alpha}
+ \sum_{i\alpha} (\epsilon_f -\mu) \hat{n}_{fi\alpha}
\nonumber \\
&+ \sum_{i\alpha} \left( V_i c_{i\alpha}^{\dag} f_{i\alpha} + {\rm h.c.} \right)
+ \frac{U}{2} \sum_{i, \alpha \neq \beta} \hat{n}_{fi\alpha} \hat{n}_{fi\beta},
\label{eq:H}
\end{align}
where $\hat{n}_{fi\alpha} = f_{i\alpha}^{\dag} f_{i\alpha}$ and the $f$ states have $N$-fold degeneracy labeled by $\alpha$~\cite{footnote-N6}.
Considering the limit $U=\infty$, we restrict the local $f$ states to the $f^0$ and $f^1$ configurations.

As described above, the hybridization strength $V_i$ is treated as a random variable distributed in accordance with the probability $P(V_i^2)$.
We expect a continuous distribution for $V_{i}^{2}$ in quasicrystals.
For simplicity, we consider a uniform distribution of width $2 \delta V^2$:
\begin{align}
P(V_i^2) = \begin{cases}
1/(2\delta V^2) & (\overline{V^2}-\delta V^{2}< V_i^2 < \overline{V^2}+\delta V^2) \\
0 & ({\rm otherwise})
\end{cases}.
\end{align}
In Eq.~(\ref{eq:H}), not only $V_{i}$ but also $\epsilon_{fi}$ may be site-dependent in quasiperiodic structures.
However, since it is the quantity $V_i^2/|\epsilon_{fi}|$ that is essential in Kondo physics, $\epsilon_{fi}$ may be fixed for the present purpose.

Let us first make a simple consideration of the consequences of the hybridization distribution.
If we consider the $f$ electron on each site independently, the site dependence of $V_i^2$ may be regarded as that of the Kondo temperature $T_{{\rm K},i}$. 
The continuous distribution of $V_i^2$ is thus read as a distribution of $T_{{\rm K},i}$.
Supposing that the lowest value of $T_{{\rm K},i}$ is so small that the ground state is practically inaccessible, unscreened $f$ moments exist in the whole temperature range, giving rise to the Curie-like divergent behavior of the low-temperature magnetic susceptibility.
However, since local susceptibilities with different values of $T_{{\rm K},i}$ should be integrated, its $T$ dependence is nontrivial.
We will derive the explicit temperature dependence of the susceptibility by numerically solving the model given by Eq.~(\ref{eq:H}).

We treat the random distribution of $V_i^2$ with the coherent potential approximation (CPA)~\cite{Yonezawa73,Elliott74} and the many-body effects
with the dynamical mean-field theory (DMFT)~\cite{Georges96}.
The CPA+DMFT scheme has been applied to a wide range of correlated models~\cite{Byczuk05,Byczuk09}.
Regarding Kondo systems, the evolution from dilute Kondo systems to coherent heavy-fermion systems was discussed~\cite{Shiina95,Mutou01,Burdin07,Grenzebach08,Otsuki10}.
These calculations correspond to the substitution of rare-earth atoms with nonmagnetic ions such as lanthanum.
In contrast, the distribution of hybridization considered here corresponds to disorder on conduction electrons.
A similar situation was discussed by Miranda {\it et al.}~\cite{Miranda96,Miranda97} in the context of Kondo disorder.

A brief description of the CPA+DMFT scheme is presented in the following.
In the CPA, we take a random average over spatial configurations of $V_i^2$ for a given probability distribution $P(V_i^2)$. 
Because the average is taken, the translational symmetry is recovered for conduction electrons.
The Green function $G_{{\rm c}\bm{k}} (i\omega)$ of conduction electrons is thus given by
\begin{align}
G_{{\rm c}\bm{k}}(i\omega) = \frac{1}{i\omega - \epsilon_{\bm{k}} + \mu - \Sigma^{\rm CPA}(i\omega) }.
\label{eq:Gc}
\end{align}
Here, $\omega$ is the fermionic Matsubara frequency.
The CPA self-energy $\Sigma^{\rm CPA}(i\omega)$ is evaluated with the help of auxiliary impurity models~\cite{Georges96}.
Since $V_i$ is now site-dependent, the impurity models are defined for each site. 
The hybridization function is given by $\Delta_i(i\omega)=V_i^2 {\cal G}_0(i\omega)$ with ${\cal G}_0(i\omega)$ being the so-called cavity Green function defined by
\begin{align}
{\cal G}_0(i\omega)^{-1} = \langle G_{{\rm c}\bm{k}}(i\omega) \rangle_{\bm{k}}^{-1} + \Sigma^{\rm CPA}(i\omega),
\label{eq:G-cav}
\end{align}
where $\langle \cdots \rangle_{\bm{k}}$ means the average over the momentum.
Together with $\epsilon_f$ and $U=\infty$, we solve the effective Anderson model and evaluate the local Green function $G_{fi}(i\omega)$ of $f$ electrons, which is site-dependent.
In our calculations, we use the hybridization-expansion solver\cite{Werner06a} of the continuous-time quantum Monte Carlo method~\cite{Rubtsov05,Gull11}.
$G_{fi}(i\omega)$ is then connected to $G_{{\rm c}\bm{k}}(i\omega)$ by the self-consistency condition
${\cal G}_0(i\omega) + {\cal G}_0(i\omega) \overline{t}(i\omega) {\cal G}_0(i\omega) = \langle G_{{\rm c}\bm{k}} (i\omega) \rangle_{\bm{k}}$. 
Here, $\overline{t}(i\omega)$ is the $t$-matrix averaged with respect to $P(V_i^2)$,
\begin{align}
\overline{t}(i\omega)
= \int d(V_i^2)\,\, P(V_i^2) \bigl[ V_i^2 G_{fi}(i\omega) \bigr]
\equiv
\left< V_i^2 G_{fi}(i\omega) \right>_{V}.
\label{eq:V-ave}
\end{align}
Combined with Eq.~(\ref{eq:G-cav}), we obtain the following formula for $\Sigma^{\rm CPA}(i\omega)$:
\begin{align}
\Sigma^{\rm CPA}(i\omega)^{-1} = \overline{t}(i\omega)^{-1} + {\cal G}_0(i\omega).
\label{eq:sigma}
\end{align}
Equations~(\ref{eq:Gc})--(\ref{eq:sigma}) are solved by numerical iteration.
The integral in Eq.~(\ref{eq:V-ave}) is evaluated using the trapezoidal rule with $N_{V}=100$ stripes.
This means that we solve the impurity model $(N_{V}+1)$ times in each iteration.

Details of our numerical calculations are given as follows.
The density of states of conduction electrons is set as constant, $\rho_c(\epsilon)= 1/2D\equiv \rho_{0}$ for $|\epsilon|<D$ for simplicity.
We fix the parameters $\epsilon_f=-0.5$ and $\overline{V^2}=0.1$ in the unit of $D=1$.
The remaining parameters are $\delta V^2$ and $T$. 
The chemical potential $\mu$ is adjusted so that the average number of electrons per site per orbital is fixed at $n/N=0.6$.
The conduction band turned out to be almost half filling, $n_{\rm c}/N \simeq 0.5$, for the parameters used in this paper.
The Kondo temperature $T_{\rm K}$ in the case with $\delta V^2=0$ is estimated as $T_{\rm K} \approx 0.19$ from the expression $T_{\rm K}=D \exp[ -|\epsilon_f|/(NV^2 \rho_0)]$.

\begin{figure}[tb]
	\begin{center}
	\includegraphics[width=\linewidth]{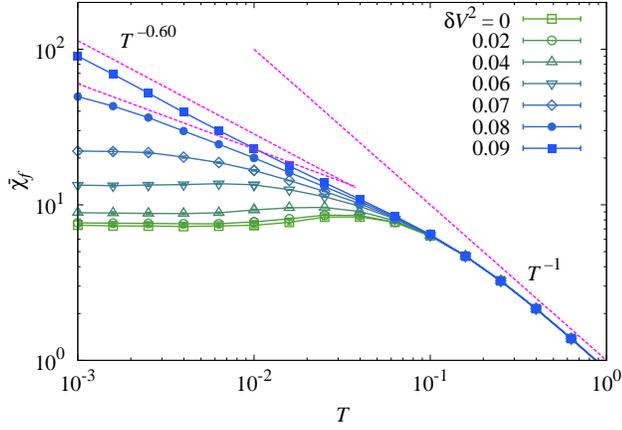}
	\end{center}
	\caption{(Color online) Temperature dependence of the magnetic susceptibility $\overline{\chi}_{f}$ for several values of the distribution width $\delta V^2$. The dashed lines indicate $T^{-1}$, $T^{-0.60}$, and $T^{-0.42}$.
	The Kondo temperature $T_{\rm K}$ is estimated as $T_{\rm K} \approx 0.19$ for $\delta V^2=0$. }
	\label{fig:chi-1}
\end{figure}

We show numerical results for the magnetic susceptibility.
The bulk susceptibility is computed by averaging the local susceptibility $\chi_{fi}$ of $f$ electrons with respect to $P(V_i^2)$,
\begin{align}
\overline{\chi}_{f} = \langle \chi_{fi} \rangle_{V}.
\end{align}
Figure~\ref{fig:chi-1} shows the $T$ dependence of $\overline{\chi}_{f}$ for several values of the distribution width, $\delta V^2$.
Here, the Curie constant of Yb$^{3+}$ ions is set to unity.
The data for $\delta V^2=0$ corresponds to the ordinary Anderson lattice model, and it shows a crossover from the Curie law $\chi \propto 1/T$ for $T \gtrsim T_{\rm K}$ to the renormalized paramagnetic Kondo state for $T \ll T_{\rm K}$.
As $\delta V^2$ increases, the low-temperature dependence changes from renormalized paramagnetic to divergent behavior.
For $\delta V^2 \gtrsim 0.08$, $\overline{\chi}_{f}$ exhibits a power-law-like behavior $\chi \sim T^{-\gamma}$ with the exponent $\gamma$ different from that in the Curie law: $\gamma \approx 0.42$ and 0.60 for $\delta V^2=0.08$ and 0.09, respectively.

\begin{figure}[b]
	\begin{center}
	\includegraphics[width=\linewidth]{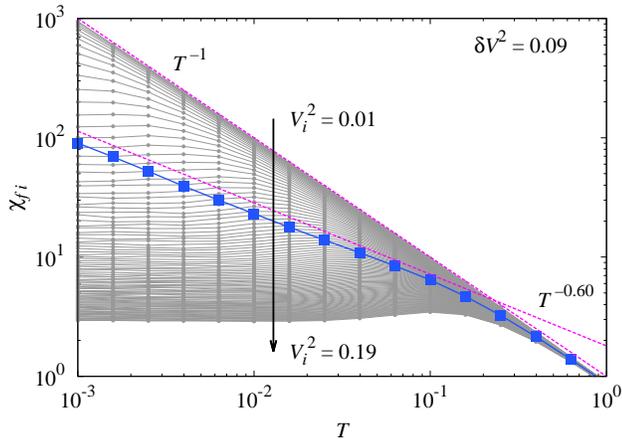}
	\end{center}
	\caption{(Color online) Temperature dependence of site-resolved magnetic susceptibilities $\chi_{fi}$ for $\delta V^2=0.09$.
	The (gray) curves from top to bottom correspond to sites with $V_i^2=0.01$ to 0.19 with interval 0.18/100.
	The squares (blue line) show the bulk magnetic susceptibility $\overline{\chi}_f$ presented in Fig.~\ref{fig:chi-1}.}
	\label{fig:chi-site-resolved}
\end{figure}

Let us discuss why the nontrivial exponents appear in the bulk magnetic susceptibility $\overline{\chi}_{f}$.
To this end, we show site-resolved susceptibilities $\chi_{fi}$ for $\delta V^2=0.09$ in Fig.~\ref{fig:chi-site-resolved}, in which there are $N_V+1=101$ lines plotted from $V_i^2=0.01$ to 0.19.
At weakly hybridizing sites, $\chi_{fi}$ follows the Curie law $1/T$ down to $T=10^{-3}$, while as $V_i^2$ increases, the Kondo behavior is recovered.
Integrating these various curves turns out to yield the power-law-like behavior with the nontrivial exponent.
As is clear from this explanation, the exponent $\gamma$ is not universal since the apparent critical behavior is not due to a critical phenomenon as in the second-order phase transition.
It is also clear that the divergence is slower than that in the Curie law, namely, $\gamma \leq 1$ in general, and $0.4 \lesssim \gamma \lesssim 0.7$ in a practical case with the flat distribution of $P(V_i^2)$.

\begin{figure}[tb]
	\begin{center}
	\includegraphics[width=\linewidth]{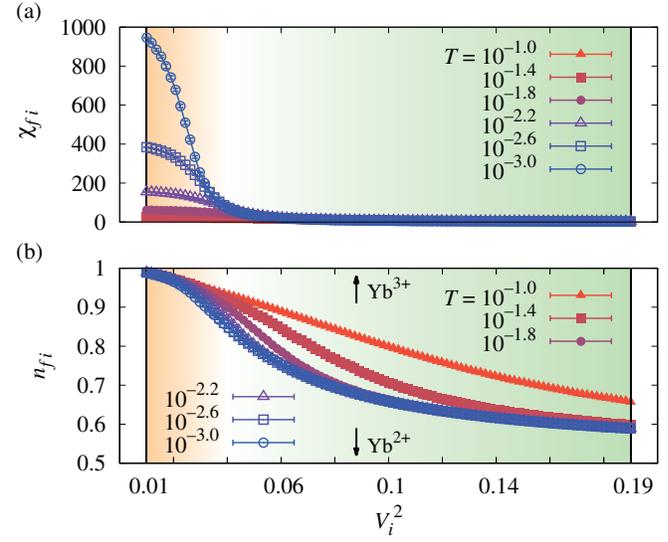}
	\end{center}
	\caption{(Color online) Site dependence of local quantities at fixed $T$: (a) magnetic susceptibility $\chi_{fi}$ and (b) $f$-electron number $n_{fi}$.
	In the hole picture, $n_{fi}=1$ ($n_{fi}=0$) corresponds to Yb$^{3+}$ (Yb$^{2+}$) ions.
	The left and right shaded areas indicate unscreened sites with a well-defined magnetic moment and intermediate-valence sites, respectively.
}
	\label{fig:V_dep}
\end{figure}

\begin{figure}[tb]
	\begin{center}
	\includegraphics[width=\linewidth]{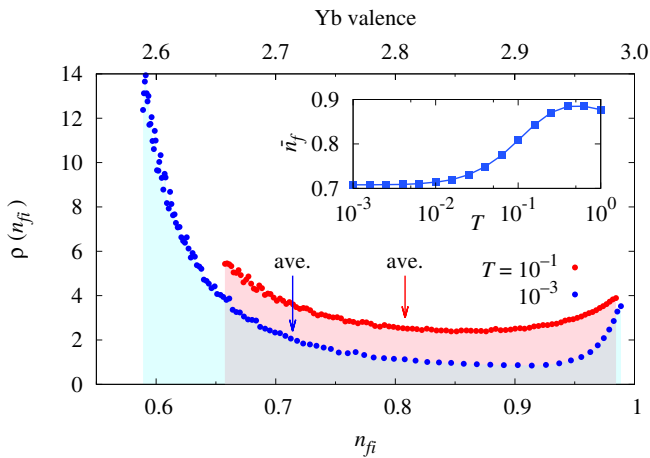}
	\end{center}
	\caption{(Color online) Distribution of the occupation number $n_{fi}$ (Yb valence $\nu_i=2+n_{fi}$) calculated from the data in Fig.~\ref{fig:V_dep}(b).
	The arrows indicate the average occupation $\overline{n}_f$ at each temperature.
	The inset shows $\overline{n}_f$ as a function of $T$. }
	\label{fig:histogram}
\end{figure}

We next present a detailed analysis of the site dependence of local quantities.
In Fig.~\ref{fig:V_dep}(a), $\chi_{fi}$ in Fig.~\ref{fig:chi-site-resolved} is replotted as a function of $V_i^2$ for several values of $T$.
A significant $T$ dependence appears in a limited region with $V_i^2 \lesssim 0.04$, indicating that only part of the sites govern the low-temperature behavior of the bulk susceptibility $\overline{\chi}_{f}$.
The number of $f$ electrons $n_{fi}$ is close to 1 (Yb$^{3+}$) at these sites as shown in Fig.~\ref{fig:V_dep}(b).
On the other hand, strongly hybridizing sites with $V_i^2 \gtrsim 0.1$ have an intermediate valence with $n_{fi} \approx 0.6$ (Yb$^{2.6+}$).
The $f$-electron valence of Yb ions is thus site-dependent.
The distribution $\rho(n_{fi})$ of the valence may be evaluated from the data for $n_{fi}$ by the formula
\begin{align}
\rho(n_{fi}) = \frac{P(V_i^2)}{ \left| dn_{fi} / d(V_i^2) \right|},
\end{align}
which is shown in Fig.~\ref{fig:histogram} for $T=10^{-1}$ and $10^{-3}$.
It turns out that the distribution $\rho(n_{fi})$ becomes considerably wider as $T$ decreases, and interestingly, it has a peak at both edges. 
The peak on the side of $n_{fi}=1$ indicates the existence of well-defined localized magnetic moments.
However, since the number of these unscreened moments is an order of magnitude smaller than that of the mixed-valence sites on the other side of the distribution, the average $f$-electron number $\overline{n}_{f}$ for the bulk inherits the mixed-valence feature as shown in the inset of Fig.~\ref{fig:histogram}.

Here, we comment on the ground state of the present model.
Although a power-law-like behavior of the magnetic susceptibility was observed, it should finally saturate at $T\to0$.
The point is that the characteristic energy scale is so small that we cannot reach the ground state in practice. 
In fact, the Kondo temperature at the site with the smallest hybridization $V_{\rm min}=0.01$ is estimated to be $T_{\rm K,min} \approx 6 \times 10^{-8}$, which is much lower than our lowest temperature of $T=10^{-3}$.

In summary, we clarified the magnetic and valence properties of Yb ions, assuming site-dependent hybridization that is randomly and continuously distributed.
When the distribution is sufficiently wide, even though most of the Yb ions are in the Kondo or mixed-valence regimes, a small number of unscreened magnetic moments exist, which make the dominant contribution to the bulk susceptibility. 
Because of the continuous distribution of $V_i^2$ as an intrinsic feature of quasicrystals, the average magnetic susceptibility exhibits a nontrivial temperature dependence having weaker power-law-like behavior than the Curie law.
The present model therefore shows both the ``quantum critical" behavior and the mixed-valence feature observed in the Yb quasicrystal.
It is important to note that in our model, (quantum) phase transitions including critical valence fluctuations are unnecessary to explain the observed peculiar magnetic properties with an intermediate valence.
Thus, the pressure effect of our scenario should be different from those based on quantum critical phenomena, since our scenario does not depend on the ``distance'' from a quantum critical point but on the {\it nature} of the distribution of the hybridization strength.
A recent experiment under pressure has shown that ``quantum critical'' behaviors are extremely robust in quasicrystals, but they are not clearly observed except in a certain pressure range in an approximant crystal\cite{Matsukawa16}, indicating the importance of the {\it continuous} probability distribution of $V_{i}^{2}$ as we discussed.

In the present scenario, the existence of unscreened magnetic moments on the weakly hybridizing sites is essential. 
Experimentally, one can confirm the validity of the scenario by a site-selective measurement, if such a measurement is possible.
Moreover, at the fundamental level of describing a model for quasicrystals, it is important to evaluate whether the appearance of the weakly hybridizing sites is intrinsic in quasiperiodic structures.
In particular, it will be interesting if almost isolated sites emerge where $f$-electron moments remain unscreened.
In this way, since the quasiperiodicity may play a hidden role in the ``quantum critical'' behavior in Yb quasicrystals, further investigations on the pressure effect and using local probes are highly desirable.

This work was supported by JSPS KAKENHI Grant Nos. 26800172, 16H01059 (J-Physics), 15K05176, and 15H05885 (J-Physics).

\bibliographystyle{jpsj}
\bibliography{JO,footnote}

\begin{thebibliography}{10}
\bibitem{footnote-Andrade}
Actually, they computed the site-dependent local density of states $\rho_{c,i}$
  of conduction electrons with fixed $V$. The hybridization strength $\Delta_i$
  is given by $\Delta_i=\pi V^2 \rho_{c,i}$.
\bibitem{footnote-N6}
We use $N=6$ throughout this paper. Actually, $N=8$ may be suitable for
  Yb$^{3+}$ ions if the crystal field splitting can be neglected. Since the
  characteristic energies are expected to be scaled roughly by $N V_i^2
  \rho_0$, our results with $N=6$ can be interpreted as those with $N=8$ by
  scaling the values of $\overline{V^2}$ and $\delta V^2$ appropriately.
\end{thebibliography}

\end{document}